# An Educational Tool for Exploring the Pumping Lemma Property for Regular Languages*

**Josue N. Rivera and Haiping Xu**
Computer and Information Science Department
University of Massachusetts Dartmouth, Dartmouth, MA, USA
Email: {josue.n.rivera, hxu}@umassd.edu

**Abstract**–*Pumping lemma has been a very difficult topic for students to understand in a theoretical computer science course due to a lack of tool support. In this paper, we present an active learning tool called MInimum PUmping length (MIPU) educational software to explore the pumping lemma property for regular languages. For a given regular language, MIPU offers three major functionalities: determining the membership of an input string, generating a list of short strings that belong to the language, and automatically calculating the minimal pumping length of the language. The software tool has been developed to provide educational assistance to students to better understand the concepts of pumping lemma and minimum pumping length, and promote active learning through hand-on practice.*

## 1  Introduction

The regular languages and finite automata are some of the most studied topics in formal language theories [1]. The notion of finite automata, introduced by McCulloch and Pitts in 1943, revolutionized the idea of what a computational model looks like, which has brought significant contributions in computer science and engineering [2]. These include but not limited to the ideas of perceptrons (predecessors to neural networks) and logic design used in the development of modern embedded systems [3]. The significant impact that finite automaton and regular languages had made in modern civilization is well-documented.

Despite the thorough studies and many existing educational materials for regular languages and finite automata, pumping lemma for regular languages has been a very difficult topic for students to understand in a theoretical computer science course. Due to a lack of tool support, students usually have insufficient practice to clearly understand the concept of pumping length and how to prove a language is not regular using pumping lemma. In this paper, we introduce an active learning tool called MInimum PUmping length (MIPU) educational software to explore the pumping lemma property for regular languages. The goal of MIPU is to serve as an active learning tool for students to understand the pumping lemma property, which is an essential concept revealing the relationship between regular languages and finite automata through its formal proof. Active learning has been defined as a high-level learning process where students are the primary actors in the process [4]. Unlike the traditional learning model where students learn new concepts through a medium such as a textbook, active learning requires students to perform hand-on tasks and learn by doing. The aim of active learning is to have students learn from experience instead of being informed about the ideas with little practical engagement. Hence, in recent years, active learning models have become a focus of discussion for teaching students in the classroom. They have been found to be effective in enhancing students' retention, boosting higher order thinking and reasoning skills, and improving student performance in STEM courses [5].

As an intriguing property of regular languages, the pumping lemma allows one to prove a language is not regular by showing the language does not satisfy the pumping lemma property. Such a proof requires one to clearly understand the concept of pumping length, how a string can be split into substrings in accordance with the property, and how it can be pumped. With MIPU, we attempt to provide three major features that contribute to the overall understanding of the pumping lemma and the concept of minimum pumping length. First, the software assists in verifying if a string belongs to a regular language described by a regular expression. By converting a regular expression into a finite automaton, we can determine if a string is a member of a given regular language. Second, the software can generate a list of short strings of a regular language. As a regular expression defines the pattern of a regular language, by generating the short strings, students can gain a better understanding of the language. Lastly, this tool can automatically calculate the minimum pumping length of a regular language and demonstrate how a given string belonging to the regular language can be split into three substrings that satisfy the pumping lemma property.

---

## 2  Related Work

With the advance of powerful personal computers and the Internet, access to educational tools become much closer within reach than in any other time in history. Gradually, educational tools have become widely available online that help to explain many advanced topics in a variety of fields. With the rise, there has been an increasing number of active learning applications that focus on aiding STEM education – a critical subject to teach in our modern lives. Computer science education is particularly crucial due to the numerous influential advancements that have emerged from the field. Thus, it is not wonder that many of these applications introduced are directed towards enhancing the experience of learning complex topics in the areas of study.

It has been proven that active learning can strengthen the experience of STEM students in the classroom [5]. A research performed by Kim and her colleagues in 2012 elaborated on and described the effects that active learning modules may have in enhancing students' critical thinking [6]. Their study had two goals: to examine the levels of critical thinking exhibited in individual reports over the semester, and to explore the effect of active learning on undergraduate students' critical thinking. With the goals in mind, they focused on designing appropriate strategies to foster innovation in an undergraduate general science course. Their team used the strategies to support students in engaging in hand-on practice by providing the learning environments that required the use of scientific knowledge in solving real-life problems. The designs included support of cognitive process such as scaffolding strategies and tools for building a knowledge pool. The modules presented to the students to evaluate critical thinking dealt with the understanding of evacuation plans for hurricanes and authentic problems associated with global warming. The study showed that the active learning strategies had been helpful to promote students' critical thinking. In recent years, there has been a push to bring effective active learning tools and strategies into the classroom to enhance the learning process of students. This trend has greatly motivated our research in developing effective tools to support active learning in computer science.

The use of educational tools in computer science classrooms has seen a significant emergence. Computer science is now an integral part in the society that we live in for the role that it plays in many crucial aspects of it. In a recent paper, Wang from the University of Toledo tackled the integration of educational tools in computer science courses [7]. He presented multiple modern software tools to assist with various subjects in a database course. He first introduced different components in a typical database course, such as Entity Relation (ER) diagram and MySQL. Then he introduced existing support tools that make the various component more interactive and easier to learn. The results of implementing these strategies in his online database course was an increase in the visual appeal of the taught contents along with a significant jump in the average grade of the class in various subjects. While his research was intended to be applied to online courses, the principles learned can be easily transferred to an in-person setting. Wang's work is an example of the shift in computer science education that is attempting to make learning more interactive and enable topics to be learned from experience rather than through passive learning.

There are currently many existing tools for experimenting with topics related to formal languages and automata, such as deterministic finite automata (DFA), nondeterministic finite automata (NFA), conversion from NFA to DFA, pushdown automata (PDA) and multi-tape Turing machines. Among the existing tools, the Java Formal Languages and Automata Package (JFLAP) is by far one of the most popular educational tools. JFLAP is a collection of graphical tools that can be used as an aid in learning the basic concepts of formal languages and automata theory [8] [9]. The goal of the tool is to "enhance the formal languages course, changing it from a traditional mathematics course into a 'hands-on' computer science course" [10]. In JFLAP, the graphical interface allows one to build automata, run them with different input strings, and see a snapshot of the automaton at any stage of the computation along with the different configurations that lead to a final state. Despite that it is a powerful tool, JFLAP lacks in some major areas of formal languages and finite automata theory, e.g., the tool support for calculation of minimum pumping length and facilitating students to understand pumping lemma property. To the best of our knowledge, there are no existing educational tools that support those features. As such, our work is complementary to other research efforts, e.g., JFLAP, that use software tools to support hand-on computer science education.

## 3  Tool Support for Pumping Lemma

Pumping lemma is a theorical idea that cannot be easily presented to students through a traditional visual medium or an intuitive explanation. Instead, it requires students to go through a sufficient number of cases to build a mental model of the concepts. Therefore, the design of an effective active learning tool for understanding pumping lemma is crucial for a successful education in theoretical computer science.

### 3.1  Pumping Lemma for Regular Languages

Aiding students in understanding pumping lemma is the core goal of MIPU. Pumping lemma is a property that all regular languages have, which can be demonstrated using a finite automaton. For this reason, it is important to understand finite automata in to learn how pumping lemma works. A regular language is defined as a set of strings that can be accepted by some finite automaton. A finite automaton is commonly seen as a computational model with a limited number of states that contain transitions between states labeled by symbols from a finite alphabet. Some or none of the states in a finite automaton are accept states and one of the states is a start state. To compute an input string, an automaton reads each symbol in the string in order and transitions to states according to the transition function. Once all symbols in the string have been processed, if a current state of the automaton is an accept state, the string is accepted; otherwise, the string is rejected. Two types of finite automata are DFA and NFA, which are equivalent. The strength of finite automata emerges from its ability to represent real-world computation using a simple model. The act of switching on and off a light is one such example, but finite automata can be used to model more complicated situations, e.g., representing the states of characters in a game or performing pattern recognition on strings.

An intuitive way of distinguishing regular languages from nonregular languages is to determine if the modeling machine needs to have an unbounded memory to account for the unlimited number of possibilities. However, this intuitive approach does not always work. For example, in the following two languages *C* and *D*, both are seemingly non-regular, but surprisingly, one of them (language *D*) is in fact regular [11].

- *C* = {*w* | *w* has an equal number of 0s and 1s}
- *D* = {*w* | *w* has an equal number of occurrences of 01 and 10 as substrings}

We can formally prove language *D* is regular by designing a regular expression that describes the language. However, one may try to design a regular expression to describe language *C*, but still fail to find one. Can we conclude *C* is not regular because no one is able to design a regular expression to describe *C*? The answer is no, and thus, it is important to establish a formal approach to assisting in determining the non-regularity of a language.

The pumping lemma for regular languages is a technique for proving non-regularity. The pumping lemma states that all regular languages have a special property, i.e., the pumping lemma property. Therefore, if a language does not demonstrate the pumping lemma property, the language must be nonregular. The pumping lemma ensures that any string in a regular language with at least a certain length, i.e., the pumping length *p*, can be "pumped" and still belong to the language. Pumping a string, in the context of the property, refers to repeating or eliminating a section of a string and still maintaining its membership with the language.

The pumping lemma can be described as follows [11], if *A* is a regular language, then there is a positive number *p* (the pumping length) where if *s* is any string in *A* with a length of at least *p*, then *s* can be divided into three substrings, *s* = *xyz*, satisfying the following three conditions:

1) for each $i \geq 0$, $xy^iz$ belongs to *A*,
2) $|y| > 0$,
3) and $|xy| \leq p$.

As demonstrated earlier, intuitively understanding the regularity and non-regularity of a language might not be sufficient. Pumping lemma has played an important role in helping understand regularity and proving a language is not regular by contradiction. However, a correct proof for nonregularity of a language requires accurate understanding of the pumping lemma for regular languages. The goal of MIPU is to aid in understanding the pumping lemma property, and based on the conditions required to satisfy the pumping lemma property, the tool provides three major functionalities: membership testing, generation of strings that belong to a regular language, and calculation of the minimum pumping length needed to demonstrate the existence of the property in a language if it is regular.

### 3.2 A Framework of the Active Learning Tool

To make MIPU easily customizable and flexible to optimize, it was built with an object-oriented design (OOD) in mind. This would enable specific components of the tool to be adjusted without affecting the overall functionality. The framework of MIPU consists of four major components that represent the major concepts in formal languages and automata. Fig. 1 showcases their corresponding classes and their interactions with each other.

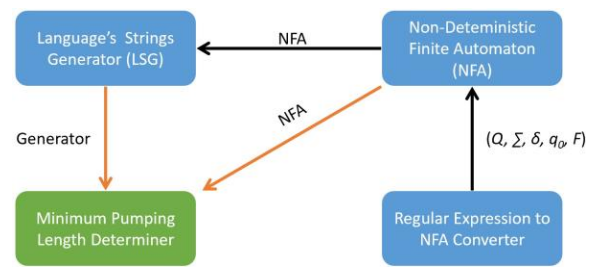

Figure 1. A framework of MIPU with four major components

As shown in Fig. 1, the four components of MIPU are a regular expression to NFA converter, an NFA simulator, a language's strings generator (LSG), and a minimum pumping length determiner. For membership testing, the regular expression to NFA converter is used to transform a given regular expression into an NFA instance that can be easily operated on. This NFA instance is bundled with a "compute" function that is used to determine if a given string is a member of the language. To generate short strings, the language's strings generator is used to generate a list of such strings that belong to the language described by the regular expression. Lastly, the determination of the minimum pumping length of a regular language uses all the components in MIPU as needed by the pumping lemma for regular language. These functionalities are further discussed in Section 4.

#### 3.2.1 Regular Expression to NFA Converter

The regular expression to NFA converter takes a regular expression in the form of a string and decodes it into a tuple of five elements that comprise an NFA. These elements include: a finite set of states (*Q*), a finite set of the alphabet that forms the language ($\Sigma$), the transition function between states ($\delta$), a start state ($q_0$), and finally, a set of accept states (*F*). Algorithm 1 shows how to generate these elements of 5-tuple. The algorithm first checks if the regular expression represents a base case, which can be an empty set, an empty string, or a regular expression containing only one symbol. Then the regular expression is parsed into a list of segments that can be iterated through to form an NFA.

**Algorithm 1:** Convert a regular expression into an NFA

**Input:** regular expression *regExp*
**Output:** *T* as 5-tuple (*states*, *alpha*, *transfun*, *startq*, *acceptq*)

1:   initialize *states* and *alpha* to empty sets
2:   initialize *transfun* to an empty map with state and symbol as key and traversable states as value
3:   *currq* = 0
4:   createNFA(*regExp*)
5:     **if** *regExp* is an empty set
6:       **return** *T* with $q_{currq}$ as start state and no accept state
7:     **else if** *regExp* is the empty string
8:       **return** *T* with $q_{currq}$ as the start and accept state
9:     **else if** *regExp* is of length 1
10:      add transition between $q_{(currq++)}$ and $q_{(currq++)}$ with *regExp* as the transition symbol
11:      add $q_{(currq-1)}$ and $q_{(currq-2)}$ to the states set
12:      add *regExp* to the alphabet set
13:      **return** *T* with $q_{(currq-2)}$ and $q_{(currq-1)}$ as the start and accept state, respectively
14:   *seg* = parseSegments(*regExp*)
15:   **for each** segment *s* in *seg*, where *s* is not an operation
16:     *T_seg* = createNFA(*s*)
17:     *start_seg*[*s*] = start state of *T_seg*
18:     *accept_seg*[*s*] = accept state of *T_seg*
19:   **for each** segment *s* in *seg*, where *s* is star
20:     update *currq* and add new states to *states* set
21:     add transitions starting with start state of the previous segment and ending with $q_{(currq+2)}$
22:   **for each** segment *s* in *seg*, where *s* is concatenation
23:     update *currq* and start & accept states
24:     add epsilon transition between the previous segment and the next segment
25:   **for each** segment *s* in *seg*, where *s* is union
26:     update *currq* and add new states to *states* set
27:     add transitions to connect the previous segment and the next segment
28:   **return** *T* with start and accept state of *seg*

For the symbol that represents the empty set, an NFA is returned with the current state (*currq*) as the start state and there is no accept state. For the empty string, an NFA is returned with *currq* as both the start and accept state. Lastly, for a regular expression that contains only one symbol other than a regular operation, two states are created (*currq*++ and *currq*++), which are connected by a transition labeled by the symbol. When the regular expression does not represent a base case, it is parsed into a list of segments. The procedure utilized to parse the expression into segments will later be discussed in Algorithm 2. The segments are iterated through in four different *for*-loops. The first *for*-loop traverses all the elements that are not an operation and perform recursive calls on Algorithm 1 for the individual segments until the base cases are reached. The following three *for*-loops are ordered according to the precedence of the regular operations, namely star, union, and concatenation. For each regular operation, the algorithm follows the standard regular expression to NFA conversion techniques [11]. New states and transitions are added as needed to the segment(s) that the operation is applied to; meanwhile, *currq* is also updated. The start and accept state of the segments involved synchs to reflect in the newly created NFA. It must be highlighted that for union and star operations, the NFA is adjusted to contain a single accept state. Fig. 2 showcases these changes. After all the segments are constructed, the 5-tuple representing an NFA is returned.

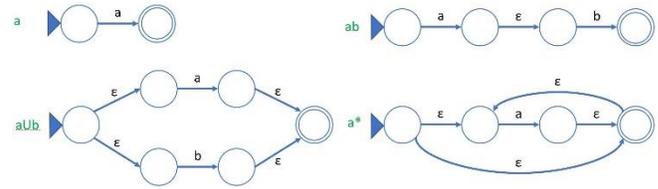

Figure 2. Conversion of regular expression to an NFA

To make the conversion procedure from a regular expression into an NFA more flexible and efficient, Algorithm 2 is used to section a regular expression into segments while building the entire NFA.

**Algorithm 2:** Parse a regular expression into a segment list

**Input:** regular expression *regExp*
**Output:** expression segment list *seg*

1:   parseSegments(*regExp*)
2:     initialize *count* to 0 and *temp* to an empty string
3:     initialize *seg* to an empty list of strings
4:     **for** *i* = 1 **to** *regExp.length*
5:       **if** *regExp.charAt*(*i*) == '('
6:         *count*++
7:         **if** *count* == 1 **continue**
8:       **else if** *regExp.charAt*(*i*) == ')'
9:         *count*--
10:       **if** *count* == 0
11:         add *temp* to *seg* and reset *temp* to an empty string
12:         **if** *i* < *regExp.length*-1 and *regExp.charAt*(*i*+1) is not star or union
13:           add concatenation operation "." to *seg*
14:         **continue**
15:     *temp* += *regExp.charAt*(*i*)
16:     **if** *count* == 0 // *temp* is an operation or one symbol
17:       add *temp* to *seg* and reset *temp* to an empty string
18:       **if** *i* < *regExp*.length-1
19:         add concatenation operation "." to *seg* if needed
20:   **return** *seg*

Algorithm 2's role is to decipher a regular expression into a list of segments that Algorithm 1 can easily convert them into an NFAs. The algorithm traverses each symbol of the regular expression, while at the same time, it keeps track of the appearance of parenthesis (*count*), the segments of the expression (*seg*) and a temporary buffer for the current segment (*temp*). For each character iterated, the character is first processed to discern parentheses. This step is performed to determine if the upcoming elements of the expression are isolated from the rest of the elements. This is essential for operations like union that requires all the elements to the right and left of the operation to be passed as inputs. If the current character is an opening parenthesis, *count* is increased by one, and the procedure immediately moves on to the next symbol.

On the other hand, if the character is a closing parenthesis, *count* is decreased by one, and the collected elements in *temp* is added into *seg* when *count* becomes zero. In addition to the elements added thus far, a concatenation operation is added as well if the next character is not a *star* or *union* operation. These components ensure that isolation is secured. If the character is not a parenthesis, it is added into *temp*. When *count* equals zero, *temp* must contain an operation or a single symbol, which is added into *seg*. In this case, a concatenation operation is added if needed. To better illustrate the functionality of Algorithm 2, a sample input and its corresponding output are provided as follows.

>   *Input = "a(caUac)c\*cac"*
>   *Output = ["a", ".", "caUac", ".", "c", "\*", ".", "c", ".", "a", ".", "c"]*

One aspect of Algorithm 1 and 2 that must be highlighted is that they require the omission of special characters as element in the NFA alphabet. The character used to represent union, concatenation, star, empty language, and epsilon cannot be elements in the alphabet. Due to this notion, the algorithms have default characters that they treat as these special symbols. Union is represented by uppercase letter "U"; concatenation is portrayed by the period "."; and the star operation is symbolized by the star character "*". The empty language is equivalent to the backslash (\), and lastly, the empty string epsilon is depicted by the lowercase letter "e". Future improvement to MIPU will allow customized settings to overwrite the default characters used.

### 3.2.2   Nondeterministic Finite Automaton (NFA)

The NFA class in the framework takes the 5-tuple generated by the regular expression to the NFA converter and offers methods for managing the NFA. One such method is to test membership of an input string. To compute the input string, the states of the NFA are traversed based on the symbols in the input string, and membership is determined if one of the possible paths leads to an accept state. This NFA model is passed to the language's strings generator and the minimum pumping length determiner for each to serve their respective roles.

The membership testing of an input string results from three individual algorithms that contribute to each other to decide if the current state ends is an accept state after a string is computed. Algorithm 3 shows this process that iterates through the character in an input string and transits to other states based on the character read. At the end of the iteration, this algorithm returns true or false depending on whether or not the current is found to be an accept state.

Algorithm 4 performs the transition method used in Algorithm 3. The algorithm searches for all possible states that the current list of states can traverse to. It will then remove those states and update the list to reflect the most recent version of the states that the current list of states has moved to. As the NFA may have multiples states that it can traverse to from the current state and an input symbol, the transit algorithm (Algorithm 4) is separated from Algorithm 3 for simplicity.

---

**Algorithm 3:** Compute a string

**Input:** *inputStr, transitions*
**Output:** *membershipStatus*

1:   initialize *current* to an empty list
2:   add start state to *current*
3:   updateEpsilonTransitions(*current, transitions*)  //Alg. 5
4:   **for each** symbol *c* in *inputStr*
5:      transitState(*c, current, transitions*)     //Alg. 4
6:      updateEpsilonTransitions(*current, transitions*)  //Alg. 5
8:   **if** *current* state is an accept state
9:      **return** true
10:  **else**
11:     **return** false

---

**Algorithm 4:** Transit between NFA states

**Input:** *symbol, current, transitions*
**Output:** *current*

1:   transitState(*symbol, current*)
2:      **if** *symbol* is epsilon
3:         **return** *current*
5:      *size* = the size of the *current* list
6:      **for** *i* = 1 **to** size
7:         **if** there is a *transition* for current state *i* and *symbol*
8:            **for each** traversable state *s* from *current* state *i*
9:               **if** state *s* is not a member of *current*
10:                 add state *s* to *current*
11:         remove state *i* from *current*
12:     **return** *current*

---

An intriguing property of the NFA is the use of a special transition called *epsilon* transition. An epsilon transition allows for the finite automaton to traverse without the need of an input symbol. The traversal of this type of transition is encapsulated in Algorithm 5. The algorithm iterates a changing list that updates within the method itself. The logic behind this approach is that if an epsilon transition is found, it is possible that the destination state may also contain another epsilon transition leading another state. However, this method has a hidden issue: if a cycle of epsilon transitions exists, this would lead to an infinite loop. The solution to this is to check if a new traversed state already exists in the list before it is added into the current list.

---

**Algorithm 5:** Update epsilon transitions

**Input:** *current, transitions*
**Output:** *current*

1:   updateEpsilonTransitions(*current, transitions*)
2:      **for each** state *i* in *current*   // *current* changes in the loop
3:         **if** there is an epsilon *transition* from *current* state *i*
4:            **for each** traversable state *s* from *current* state *i*
5:               **if** *state s* is not a member of *current*
6:                  add state *s* to *current*
7:      **return** *current*

---

The algorithms presented form the bases for the membership testing functionality of MIPU. After traversing the NFA graph and tracking all possible paths, one can determine the membership of a string by observing if one of the paths leads to an accept state. The ability to detect the membership of a string is essential for the next two components of the MIPU framework, namely the language's strings generator and minimum pumping length determiner.

### 3.2.3     Language's Strings Generator (LSG)

The language's strings generator uses a given NFA instance to generate an adjustable number of permutations from the alphabet. These permutations must be strings that can be accepted by the finite automaton. Every so often, the generator generates a new batch of strings and stores them in a buffer for future usage. To improve the performance of the permutation process for strings, branches of a permutation tree are tracked. If a path will not likely lead to a final state along the way, that branch is removed. The fate of a future branch can be determined by observing the current states that the NFA is tracking for the current segment of the string that has been generated thus far.

### 3.2.4     Minimum Pumping Length Determiner

Finally, as one of the primary functionalities of MIPU, the minimum pumping length determiner can calculate the minimum pumping length of a regular language according to the definition of pumping lemma. The tool also retrieves one of the shortest strings in the language that meet the conditions and partitions it into three segments $x$, $y$, and $z$ described in pumping lemma. The method takes an NFA instance and the strings generated by the LSG as inputs and tests the conditions to derive how the pumping lemma property is satisfied. Since the strings are ordered by their string lengths, we will be able to check strings starting from the shortest one and determine the minimum pumping length that meets the pumping lemma requirements.

## 4    Pumping Lemma for Regular Language

The pumping lemma presents a set of conditions that must be satisfied in order to demonstrate the pumping lemma property. These conditions include testing the membership of a "pumped" string, where the original string belongs to a regular language and is of a size greater than or equal to the minimum pumping length. To help with the correct understanding of the pumping lemma concept, MIPU offers three main tools that are essential to determine the existence of the property in regular language, which are membership testing, string generation, and automated minimum pumping length determination, as illustrated in Fig. 3. Membership testing function determines if an input string is a member of a given regular language, which can be used to verify if a string still maintains its membership with the language after being pumped. String generation is the retrieval of an ordered list of strings that belong to the language. This functionality is critical for validating that a significant number of strings in the language adheres to the conditions set by the pumping lemma. Lastly, as the name suggests, the minimum pumping length determiner automatically calculates the minimum pumping length of a regular language described by a regular expression. It also, along with the minimum pumping length, provides the short strings that meet the conditions of the pumping lemma and the ways how the strings can be partitioned into three appropriate substrings $x$, $y$ and $z$. These are core concepts that encompass the tools needed to determine the non-regularity of certain language using pumping lemma.

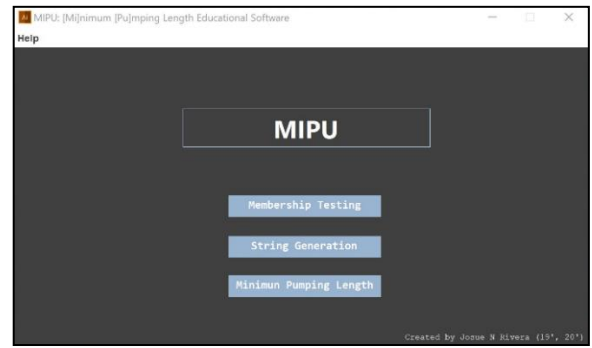

Figure 3. Main menu of MIPU

### 4.1   Membership Testing for Regular Languages

The membership testing module is composed of the regular expression to NFA converter and the NFA class described in Sections 3.2.1 and 3.2.2, respectively. The core of the functionality is found in the "compute" method of the NFA class. The method traverses a graph created during the conversion of the regular expression to an NFA and observes if there is a path leading to an accept state.

As shown in Fig. 4, MIPU allows one to enter a regular expression and an input string. Then it takes the regular expression and generates an NFA for it. While computing membership, the input string is passed as a parameter to the NFA's "compute" function, which returns either "True" or "False", indicating whether the sting belongs to the language or not.

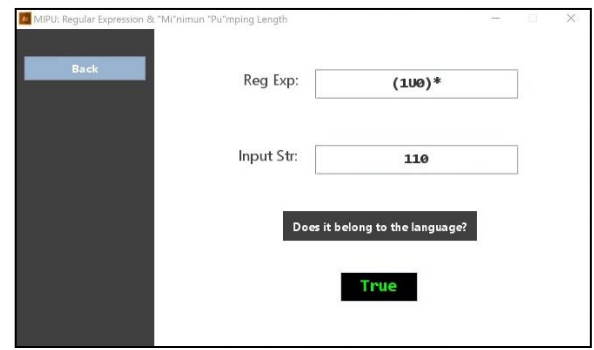

Figure 4. Membership testing window after a string is tested

Fig. 5 presents another example for membership testing, where the regular expression is (1∪0)*101(1∪0)* and the input string is 1011. As the result shows, the input string is determined to be a member of the language. The substring 101 of the given string reflects the segment 101 of the regular expression, while the symbol "1" at the end of the input string is the one generated by the rightmost segment (1∪0)*. Due to the tool's ability to track multiple paths of the NFA as it computes a string, the only path that leads to an accept state for 1011 can be identified to accept the string.

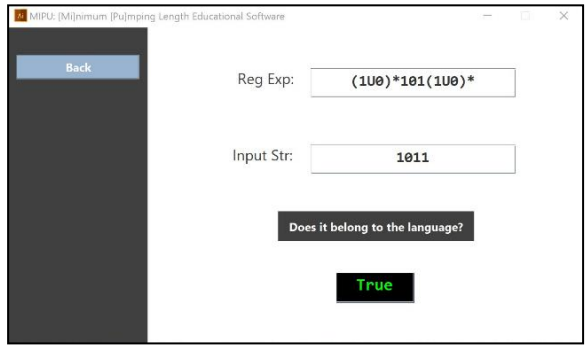

Figure 5. Another example for membership testing

### 4.2 String Generation

String generation for a given a regular expression is the second tool offered by MIPU. It is responsible for producing strings that are members of the regular language. The resulting strings are ordered by the length of the strings from the shortest to the longest. The generator can dynamically generate more strings as requested. This functionality uses the following components: regular expression to NFA converter, the NFA class, and the LSG. The LSG module uses the NFA produced from the regular expression and generates the strings from permutations of its alphabet that are members of the language. Various optimizations are used to eliminate branches of a permutation that will not lead to a valid string.

The string generation tool allows a user to enter a regular expression in the provided text field. After the regular expression is converted into an NFA, an LSG instance is created to generate strings that are recognized by the NFA. The LSG module dynamically calls a "generate" function that produces new strings as requested. Fig. 6 shows some resulting strings after the "Get Strings" button is pressed. The generated strings belonging to the language are listed in a lexicographic order, which is the same as the dictionary ordering except that shorter strings precede longer ones.

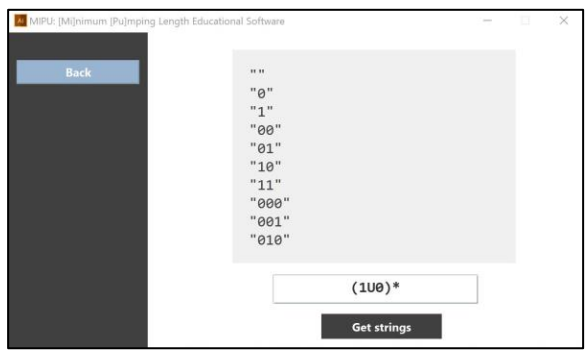

Figure 6. An example of generating short strings

Fig. 7 shows another example for string generation. Note that the shortest string "00" is generated first by ignoring the segments containing a star operation. Then the following strings are generated by considering the segments containing a star operation, e.g., the last "1*" segment.

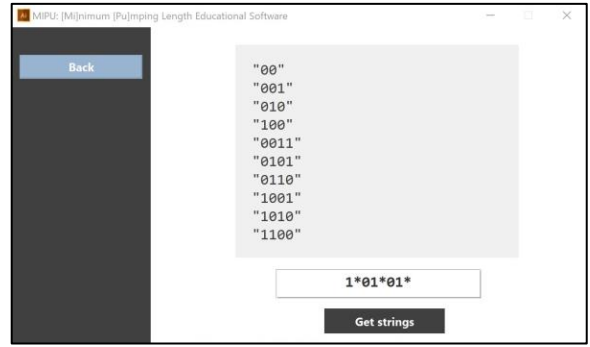

Figure 7. Another example for string generation

### 4.3 Determination of Minimum Pumping Length

The last function implemented in MIPU is to automatically calculate the minimum pumping length of a regular language. All modules of the MIPU framework, including conversion of a regular expression into an NFA and testing the various pumping lemma conditions, are used to achieve this function. As shown in Fig. 8, the minimum pumping length determination tool requires only a regular expression as its input. Once a regular expression is put in, an instance of the minimum pumping lemma determiner is created, which tests a significant number of strings belonging to the language and then decides the minimum pumping length. The figure shows that when the regular expression "10*1" is typed in and the "Get Min Pump" button is pressed, the tool displays the minimum pumping length of the regular language along with a string example "101" that helps explain a way of portioning of the string that satisfies the pumping lemma conditions.

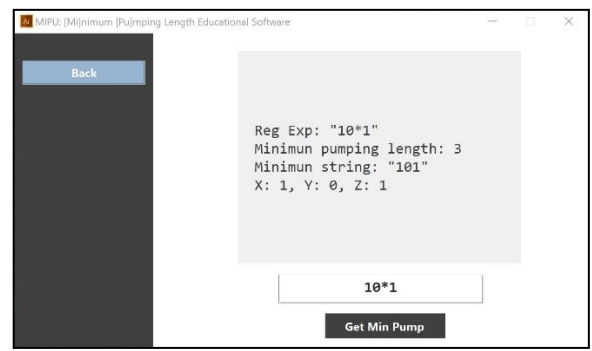

Figure 8. Minimum pumping length determination

Fig. 9 shows the minimum pumping length of the regular language 1*01*01. In this scenario, the minimum pumping length is 3 and one of the minimum strings that meets the conditions of the pumping lemma property is 001. A possible partition of the string is also displayed. It should be noted that although 001 is selected, other minimum strings also exist, e.g., 100 and 010. One aspect of the results produced that should also be highlighted is the minimum string 001 given in Fig. 9 in comparison to the shortest string 00 shown in Fig. 7. In both scenarios, the regular expressions are the same, but the shortest string generated in Fig. 7 cannot be pumped, thus it is not listed as a minimum string.

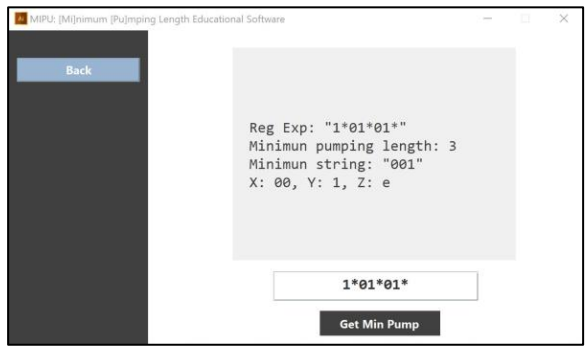

Figure 9. Another example of minimum pumping length

One last example of minimum pumping length, illustrated in Fig. 10, is the regular expression *aab*U*a\*b\**. The result is interesting because normally with a union operation where the left segment of the union operation represents a finite language and the right segment represents an infinite language, the minimum pumping length would be larger than the length of the finite segment since the string represented by the finite segment usually cannot be pumped. However, in this particular example, because the left segment can be generated by the right segment, the minimum pumping length of the regular expression equals to the minimum pumping length of the right segment, which is 1.

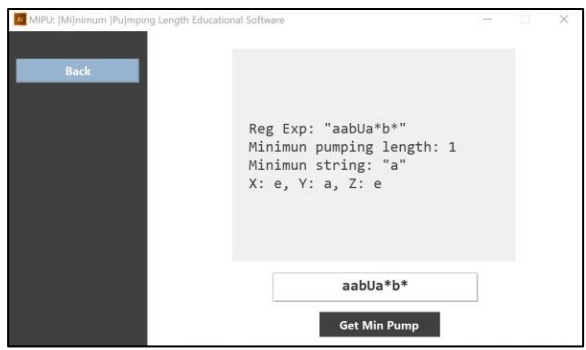

Figure 10. One more example of minimum pumping length

For more examples, the MIPU as well as the source code can be downloaded from the GitHub repository at https://github.com/JosueCom/MIPU.

## 5   Conclusions and Future Work

Finite automata and regular languages have brought humanity to a new age of innovation. They have led to advancements in artificial intelligence, the design of modern computers, and the representation of complex systems by a machine with limited memory. Through the MIPU project as well as the forthcoming improvements to enhance active learning, students will become more familiar with the formal concept of pumping lemma and overcome the complex challenge of understanding the concepts of regularity and nonregularity of languages. MIPU creates an environment that enables students to be actors for developing higher order thinking, and has the potential to be an effective tool in aiding students to better understand complex concepts.

For future work, we will improve MIPU to support visualization of the process of creating an NFA from a regular expression. We will also provide a pumping operation function that can retrieve a string that has been pumped for a given number of times. Additionally, the tool will allow a user to configure settings including redefining the restricted characters used to represent special symbols in a regular expression. The performance of generating strings may also be improved by designing a new generator that traverses the NFA graph when forming new strings instead of creating a permutation tree. Finally, we will redesign the GUI for string generation to allow dynamic generation of new strings when requested by users.